\documentclass[11pt]{article}
\usepackage{a4}
\usepackage{amsfonts}
\usepackage{amsmath,amssymb}

\usepackage{graphicx}

\makeatother

\def\un#1{\relax\ifmmode\@@underline#1\else
        $\@@underline{\hbox{#1}}$\relax\fi}

\let\du=\du                     

\def\d{\delta}

\def\f{\phi}

\def\h{\eta}

\def\k{\kappa}
\def\l{\lambda}
\def\m{\mu}
\def\n{\nu}

\def\p{\pi}

\def\D{\Delta}

\def\ve{\varepsilon}

\def\bo{{\raise-.3ex\hbox{\large$\Box$}}}               
\def\pa{\partial}                                       
\def\TH{{\raise.2ex\hbox{$\displaystyle \bigodot$}\mskip-4.7mu \llap H \;}}
\def\face{{\raise.2ex\hbox{$\displaystyle \bigodot$}\mskip-2.2mu \llap {$\ddot
        \smile$}}}                                      

\def\abs#1{\left| #1\right|}                    
\def\leftrightarrowfill{$\mathsurround=0pt \mathord\leftarrow \mkern-6mu
        \cleaders\hbox{$\mkern-2mu \mathord- \mkern-2mu$}\hfill
        \mkern-6mu \mathord\rightarrow$}
\def\dvec#1{\vbox{\ialign{##\crcr
        \leftrightarrowfill\crcr\noalign{\kern-1pt\nointerlineskip}
        $\hfil\displaystyle{#1}\hfil$\crcr}}}           
\def\dt#1{{\buildrel {\hbox{\LARGE .}} \over {#1}}}     

\def\frac#1#2{{\textstyle{#1\over\vphantom2\smash{\raise.20ex
        \hbox{$\scriptstyle{#2}$}}}}}                   
\def\sfrac#1#2{{\vphantom1\smash{\lower.5ex\hbox{\small$#1$}}\over
        \vphantom1\smash{\raise.4ex\hbox{\small$#2$}}}} 
\def\bfrac#1#2{{\vphantom1\smash{\lower.5ex\hbox{$#1$}}\over
        \vphantom1\smash{\raise.3ex\hbox{$#2$}}}}       
\def\afrac#1#2{{\vphantom1\smash{\lower.5ex\hbox{$#1$}}\over#2}}    

\def\[{\lfloor{\hskip 0.35pt}\!\!\!\lceil}
\def\]{\rfloor{\hskip 0.35pt}\!\!\!\rceil}

\def\du#1#2{_{#1}{}^{#2}}

\def\un{\underline}
\def\fracmm#1#2{{{#1}\over{#2}}}

\def\low#1{{\raise -3pt\hbox{${\hskip 0.75pt}\!_{#1}$}}}

\def\Dot#1{\buildrel{_{_{\hskip 0.01in}\bullet}}\over{#1}}
\def\dt#1{\Dot{#1}}

\def\DDot#1{\buildrel{_{_{\hskip 0.01in}\bullet\bullet}}\over{#1}}
\def\ddt#1{\DDot{#1}}

\newskip\humongous \humongous=0pt plus 1000pt minus 1000pt
\def\caja{\mathsurround=0pt}
\def\eqalign#1{\,\vcenter{\openup2\jot \caja
        \ialign{\strut \hfil$\displaystyle{##}$&$
        \displaystyle{{}##}$\hfil\crcr#1\crcr}}\,}
\newif\ifdtup

\newcommand{\be}{\begin{equation}}
\newcommand{\ee}{\end{equation}}
\newcommand{\nbe}{\begin{equation*}}
\newcommand{\nee}{\end{equation*}}

\newcommand{\lb}{\label}

\begin{document}

\thispagestyle{empty}

{\hbox to\hsize{
\vbox{\noindent April 2010 \hfill version 2 }}}

\noindent
\vskip2.0cm
\begin{center}

{\large\bf SLOW-ROLL INFLATION IN $(R+R^4)$ GRAVITY~\footnote{
Supported in part by the Japanese Society for Promotion of Science (JSPS)}}
\vglue.3in

Sho Kaneda${}^a$, Sergei V. Ketov~${}^{a,b}$, Natsuki Watanabe${}^a$ 
\vglue.1in

${}^a$ {\it Department of Physics, Tokyo Metropolitan University, Japan}\\
${}^b$ {\it IPMU, University of Tokyo, Japan}
\vglue.1in
kaneda-sho@ed.tmu.ac.jp, ketov@phys.metro-u.ac.jp, 
watanabe-natsuki1@ed.tmu.ac.jp
\end{center}

\vglue.3in

\begin{center}
{\Large\bf Abstract}
\end{center}
\vglue.1in 

\noindent We reconsider the toy-model of topological inflation, based on the
$R^4$-modified gravity. By using its equivalence to the certain scalar-tensor 
gravity model in four spacetime dimensions, we compute the inflaton 
scalar potential and investigate a possibility of inflation. We confirm the 
existence of the slow-roll inflation with an exit. However, the model suffers 
{}from the $\h$-problem that gives rise to the unacceptable value of the 
spectral index $n_s$ of scalar perturbations.

\newpage
\section{Introduction}

The identity of the inflaton field remains to be one of the main unsolved
problems in theoretical explanation of inflation \cite{inf}. The inflaton field
 should not be an {\it ad hoc} degree of freedom, but the field whose presence 
is natural from the viewpoint of gravity and/or high-energy physics. During the
 short inflationary period in early Universe the energy density was dominated 
by vacuum energy.  Inflaton coupling to a (non-gravitational) matter is 
needed to guarantee the conversion of the vacuum energy into radiation at  the 
end of inflation, whereas weakness of that coupling is needed to ensure 
 stability of the inflaton scalar potential against quantum corrections. 
Quantum fluctuations of the inflaton field are responsible for the generation
of structure in our Universe, while the fluctuation spectrum is nearly
conformal \cite{llbook}. The primordial spectrum of scalar perturbations in the
 power-law approximation takes the form of $k^{n_s-1}$ in terms of the comoving
 wave number $k$ and the spectral index $n_s$ close to $1$.

One of the natural possibilities is the identification of inflaton  with the 
self-interacting conformal mode of a metric. It can be realized in the
 modified gravity theories whose effective Lagrangian is given by a function of
 the scalar curvature. The first models of that type were proposed by 
Starobinsky in 1980 \cite{star}. Already the simplest Starobinsky model, based 
on the $(R+R^2)$ modified  gravity, is the excellent example of chaotic 
inflation, whose predictions for the spectral indices \cite{mchi} well agree 
with the recent cosmological observations \cite{wmap5}.~\footnote{See 
ref.~\cite{kkw1} for a recent discussion of the $(R+R^2)$-gravity and 
inflation.}

The modified gravity models of inflation are {\it minimal} in the sense 
that they do not rely of the Riemann and Ricci curvature dependence of the 
effective action. The Weyl tensor of a FLRW universe vanishes, so 
that the FLRW Riemann curvature can be expressed in terms of the FLRW Ricci 
tensor and the scalar curvature. A dependence upon the Ricci tensor would give 
rise to a propagating spin-2 degree of freedom, in addition to the metric 
\cite{mag}. Hence, the modified gravity models, depending upon the scalar 
curvature only, should play the most important role in cosmological dynamics.

The modified gravity theories are popular in modern theoretical cosmology 
--- see eg., refs.~\cite{sot} for recent reviews --- however, they are 
phenomenological, and are not derived from any fundamental theory of 
gravity. The most promising candidate for Quantum Theory of Gravity is given by
Superstring Theory \cite{polch}. It is unknown how to embed a modified
 gravity model into the Superstring Theory. However, there exist some 
model-independent restrictions coming from closed superstrings (it is the 
closed or type-II superstrings that are directly related to gravity). One of 
them is the absence of the quadratic and cubic curvature terms in the 
low-energy superstring effective action --- see eg., ref.~\cite{ket} for 
details, and ref.~\cite{iihk} for a recent discussion  in the context of 
inflation.~\footnote{The quadratic curvature terms may still be generated
 via superstring compactification.} In this paper we would like 
to concentrate on the inflationary model of gravity with the {\it quartic} 
scalar curvature term. The leading quantum gravitational corrections to the 
type-II superstrings are quartic indeed \cite{polch,ket,iihk}. As 
regards superstring moduli, we assume that they are stabilized by fluxes via 
compactification \cite{flux}.

The inflationary mechanism in the $(R+R^4)$ gravity is very different from that
 of the $(R+R^2)$ gravity, and is close to a {\it topological} inflation 
\cite{topo}.  The $R^4$ topological inflation was investigated earlier in 
ref.~\cite{kal} from the higher-dimensional viewpoint, namely, in five 
spacetime dimensions. Our analysis in this paper is truly four-dimensional, 
while our quantitative results in four dimensions differ from those of 
ref.~\cite{kal}. 

Our paper is organized as follows. In Sec.~2 we introduce our model of modified
gravity, establish its equivalence to the certain scalar-tensor gravity model, 
and compute the inflaton scalar potential. In Sec.~3 we briefly review
 the topological inflation \cite{topo}. Our main results are given in Sec.~4
where we describe slow-roll inflation in the $R^4$-modified gravity and compare
our results with topological inflation. Sec.~5 is our conclusion.  

\section{The model and our setup}

There is {\it a priori} no reason of restricting the gravitational Lagrangian 
to the standard Einstein-Hilbert term that is linear in the scalar curvature, 
as long as it does not contradict an experiment. Nowadays, there is no doubt 
that the extra terms of the higher-order in the curvature should appear in the 
gravitational effective action of {\it any\/} Quantum Theory of Gravity, while 
they do appear in String Theory \cite{polch,ket}. Since the scale of inflation 
in the early Universe is just a few orders less than the Planck scale 
\cite{llbook}, it is conceivable that the higher-order gravitational terms may 
be instrumental for inflation. It is already the case in the simplest modified 
gravity model with the $R^2$ terms \cite{star}. Here we consider the modified 
gravity models in four space-time dimensions, with an action 
\be \lb{fR} S_f = -\fracmm{1}{2\k^2}\int d^4x\, f(R) \ee
whose Lagrangian is given by
\be \lb{ourc}
f(R) = R -\fracmm{1}{M^{4p-2}}R^{2p}~,\qquad{\rm where}\qquad
 p=1,2,3,\ldots \ee
by paying special attention to the $p=2$ case and beyond. The parameter $M$ in 
eq.~(\ref{ourc}) has the dimension of mass.  We use the spacetime signature 
$(+,-,-,-)$ and the units $\hbar=c=1$. The Einstein-Hilbert term in our 
eqs.~(\ref{fR}) and (\ref{ourc}) has the standard normalization with 
$\k=M_{\rm Pl}^{-1}$ in terms of the {\it reduced} Planck mass 
$M_{\rm Pl}^{-2}=8\p G_N$. The rest of our notation for space-time (Riemann) 
geometry is the same as in ref.~\cite{landau}.    

In contrast to General Relativity having $f'(R)=const.$, the field $A=f'(R)$  
is {\it dynamical} in the modified gravity with $f'(R)\neq const.$ In terms of 
the fields $(g_{\m\n},A)$ the equations of motion associated with the action 
(\ref{fR}) are of the 2nd order in the derivatives of the fields. 

The new field degree of freedom represents the propagating conformal mode of 
the metric. It can be easily seen in the context of the known classical 
equivalence between the $f(R)$ gravity and the scalar-tensor gravity, via the
Legendre-Weyl transform \cite{oldr}. The equivalent action reads
\be \lb{stgr}
S_{\f} =  \int d^4x\, \sqrt{-g}\left\{ \fracmm{-R}{2\k^2}
+\fracmm{1}{2}g^{\m\n}\pa_{\m}\f\pa_{\n}\f - V(\f) \right\} \ee
in terms of the physical (and canonically normalized) scalar field $\f(x)$
having the scalar potential
\be \lb{poten}
V(\f) = -\fracmm{M^2_{\rm Pl}}{2}\exp \left\{
 \fracmm{-4\f}{M_{\rm Pl}\sqrt{6}}\right\}
Z\left( \exp \left[ \fracmm{2\f}{M_{\rm Pl}\sqrt{6}} \right] \right)
\ee
where the function $Z$ is related to the function $f$ via the Legendre 
transformation
\be \lb{clt} R=Z'(A) \qquad{\rm and}\qquad f(R)=RA(R)-Z(A(R)) \ee
whereas the arguments $A$ and $\f$ are related by~\footnote{When
 compared to the notation in  our recent paper \cite{kkw1}, we have changed 
$\f\to -\f$ here.}
\be \lb{ch1}
A(\f) = \exp \left[ \fracmm{2\k\f(x)}{\sqrt{6}} \right]  \ee

As regards cosmological applications, it is desirable to explicitly derive the
scalar potential $V(\f)$. It amounts to inverting the relation
\be \lb{invert}   f'(R) = A(\f) \ee 
that is the consequence of eq.~(\ref{clt}). In our case (\ref{ourc}) we find 
\be  \lb{spot}  
V(\f) = V_0  \exp \left( \fracmm{-2\sqrt{2}\f}{\sqrt{3}M_{\rm Pl}}\right)
\left\{ 1- \exp \left[ \sqrt{\fracmm{2}{3}}\fracmm{\f}{M_{\rm Pl}}\right]
\right\}^{\fracmm{2p}{2p-1}}
\ee
whose overall normalization constant is given by
\be \lb{const}
V_0= \left( \fracmm{2p-1}{4p} \right) M^2_{\rm Pl} M^2 \left(
\fracmm{1}{2p}\right)^{\fracmm{1}{2p-1}}
\ee
In terms of the new dimensionless variable 
\be \lb{newvar}
y=\sqrt{\fracmm{2}{3}}\fracmm{\f}{M_{\rm Pl}}
\ee
the potential (\ref{spot}) takes the form
\be \lb{spot2}
\fracmm{V(y)}{V_0} =  e^{-2y}\left( e^y-1\right)^{\fracmm{2p}{2p-1}}\equiv 
g_{2p}(y)
\ee 
This potential is bounded from below, has a local minimum at $\f_{\rm min}=0$ 
and a local maximum at 
\be \lb{max}
 \f_{\rm max}=\sqrt{\fracmm{3}{2}}\,M_{\rm Pl} \ln\left[ \fracmm{2p-1}{p-1}
\right]
\ee
In particular, when $p=2$ we find
\be \lb{pot4}
\fracmm{V(y)}{V_0}= e^{-2y}\left(e^y-1\right)^{4/3}\equiv g(y)=
\abs{z(1-z)(1+z)}^{4/3}
\ee
where
\be \lb{norm4}
V_0= \fracmm{3}{8\cdot 2^{2/3}} M^2_{\rm Pl}M^2 \qquad {\rm and}\qquad
z=\exp\left( -\fracmm{y}{2} \right) 
\ee

\begin{figure}[t]
\centering
\includegraphics[width=5cm,clip]{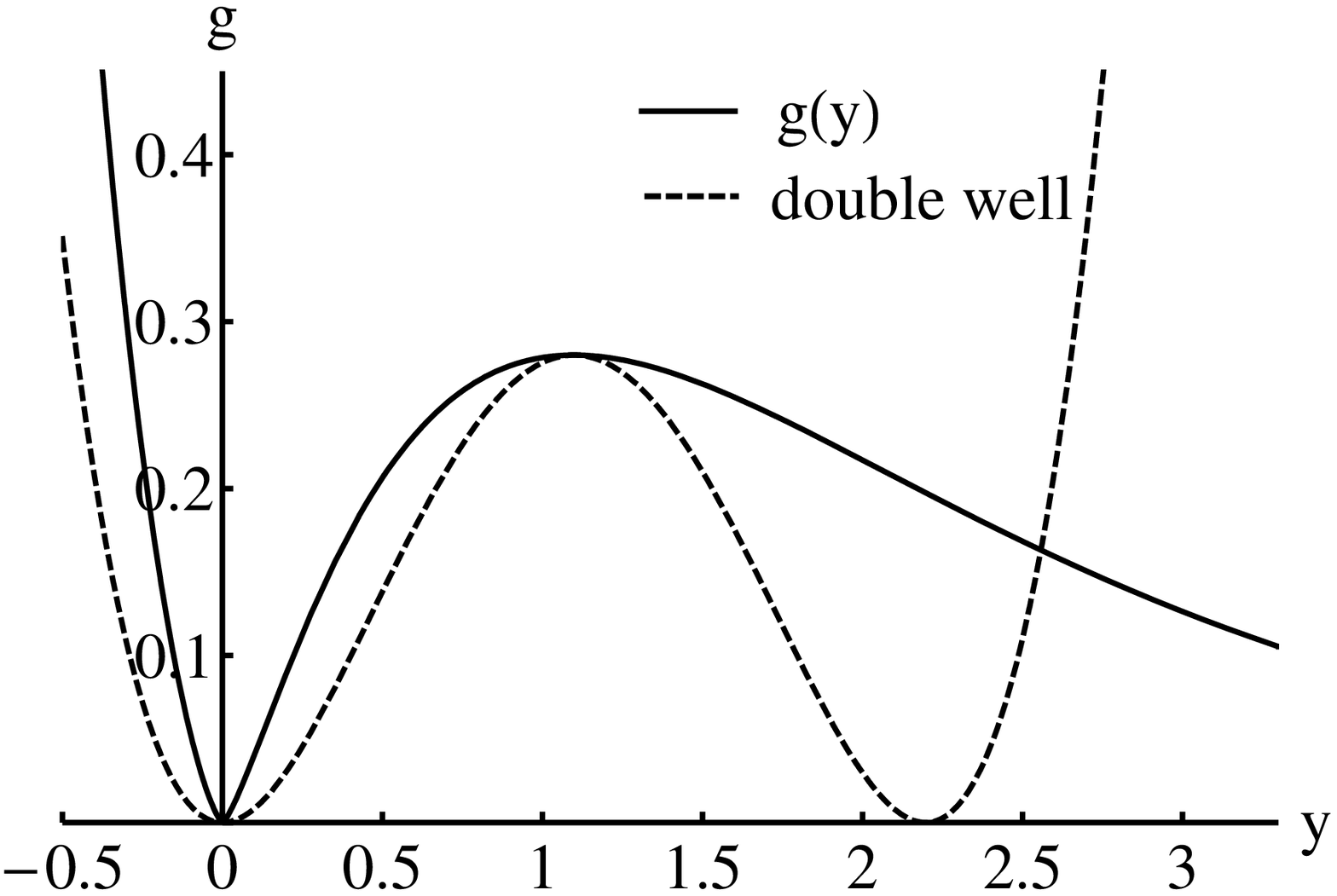}
\caption{\small Graph of the function 
$g(y)= g_4(y)=e^{-2y}\left(e^y-1\right)^{4/3}$}
\label{fig:1}
\end{figure}

\begin{figure}[b]
\centering
\includegraphics[width=5cm,clip]{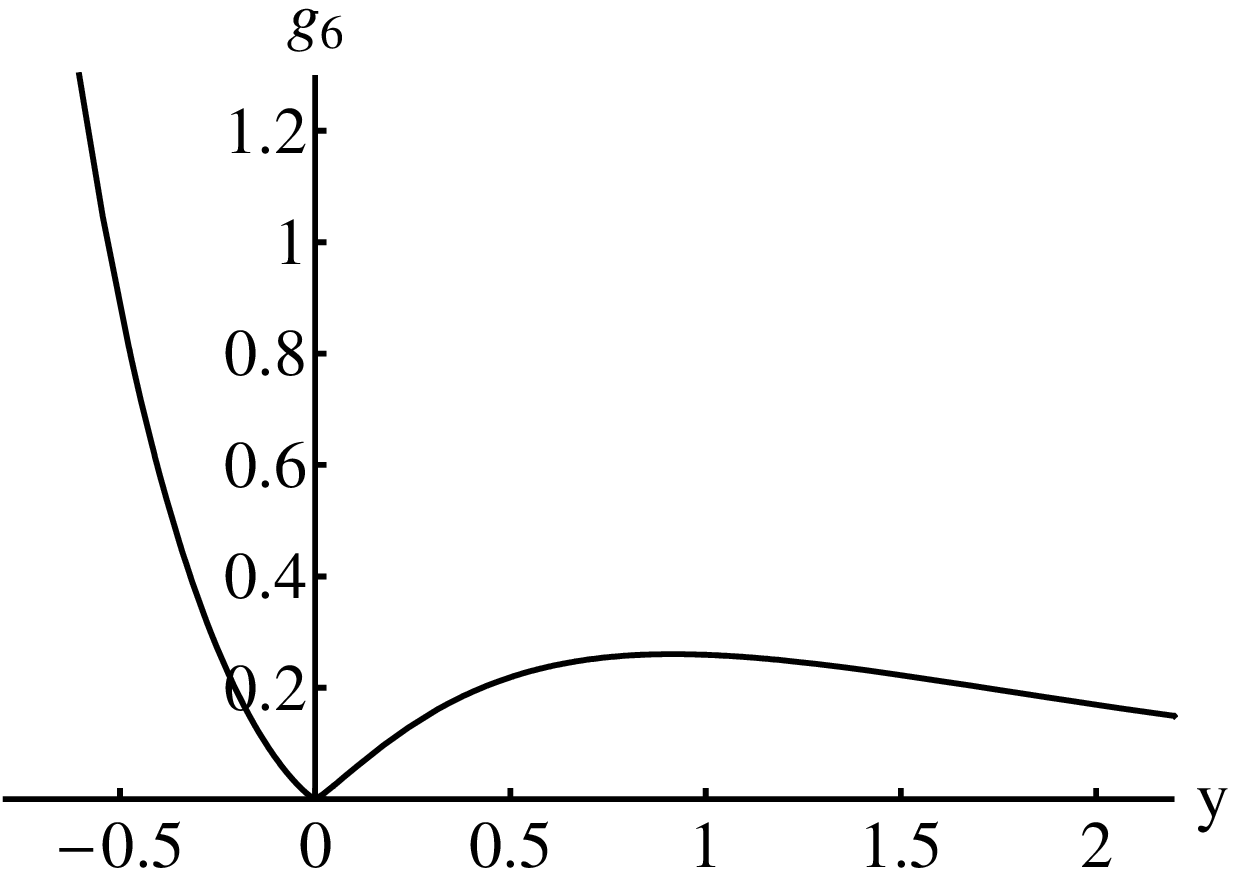}
\caption{\small Graph of the function 
$g_6(y)=e^{-2y}\left(e^y-1\right)^{6/5}$}
\label{fig:2}
\end{figure}

The graphs of the functions $g(y)\equiv g_4(y)$ and $g_6(y)$ are
given in Figs.~1 and 2, respectively. The graphs of $g_{2p}(y)$ of $p\geq 4$ 
are very close to that in Fig.~2.

In the limit $p=+\infty$ the scalar potential $V_{\infty}(y)/V_0$ (after
renormalization) reads  
\be \lb{infin}
 g_{\infty}(y) =e^{-y}\abs{1-e^{-y}}
\ee
The values of $\f_{\rm max}$ for various $p$ are given by
\be \lb{maxv}
\eqalign{
p=2: & ~~~\f_{\rm max}= \left(\sqrt{\fracmm{3}{2}}\ln 3\right)M_{\rm Pl}\approx
 1.3455~M_{\rm Pl}~, \cr
p=3: & ~~~\f_{\rm max} \approx 1.122~M_{\rm Pl}~,\quad
p=4: ~~~\f_{\rm max} \approx 1.037~M_{\rm Pl}~,\cr 
p=\infty: & ~~~\f_{\rm max}= \left(\sqrt{\fracmm{3}{2}}\ln 2\right)M_{\rm Pl}
\approx 0.85~M_{\rm Pl}~.\cr}
\ee

The $p=2$ function $g(y)$ of eq.~(\ref{pot4}) asymptotically approaches zero
as  $\exp(-2y/3)$ for $y\to +\infty$, and diverges as $\exp(2\abs{y})$ for 
$y\to -\infty$.

\section{Topological inflation}

Topological inflation was suggested by Vilenkin and Linde in 1994 \cite{topo}.
When the inflaton scalar potential has a discrete symmetry $Z_2$ (say, against 
$\tilde{\f}\to -\tilde{\f}$) and the state $\tilde{\f}=0$ is unstable, the 
Universe gets divided into different domains related by domain walls. Those 
domain walls inflate with time, while their thickness exponentially grows. 
Hence, the domain walls may appear as the seeds of inflation. The topological 
inflation does not require fine-tuning of the initial conditions \cite{topo}.

The simplest realization of topological inflation is described by an action
\be \lb{stopo}
S[\tilde{\f}] =  \int d^4x\, \sqrt{-g}\left\{ \fracmm{-R}{2\k^2}
+\fracmm{1}{2}g^{\m\n}\pa_{\m}\tilde{\f}\pa_{\n}\tilde{\f} - 
V_{\rm dw}(\tilde{\f}) \right\} 
\ee
with the Higgs-like (or double-well) scalar potential \cite{topo}
\be \lb{dw}
V_{\rm dw}(\tilde{\f}) = \fracmm{\l}{4}\left(\tilde{\f}^2-\h^2\right)^2
\ee
The $Z_2$ symmetry breaking in this model results in the formation of two
domains with $\tilde{\f}=\pm \eta$. Those domains are divided by the domain 
wall interpolating between the two minima. A static domain wall solution 
without gravity (say, in the $yz$ plane) is obtained from the Bogomol'nyi 
decomposition of the static energy (per unit area)
\be \lb{bog}
\eqalign{
E[\tilde{\f}] = & ~\int^{+\infty}_{-\infty}dx\left[ \fracmm{1}{2}
\left(\fracmm{d\tilde{\f}}{dx}\right)^2 +V(\tilde{\f})\right] \cr
& ~=\fracmm{1}{2}\int^{+\infty}_{-\infty}dx
\left( \fracmm{d\tilde{\f}}{dx} \mp\sqrt{2V(\tilde{\f})}\right)^2 
\pm \int^{\tilde{\f}(\infty)}_{\tilde{\f}(-\infty)} \sqrt{V(\tilde{\f})}
\, d\tilde{\f}\cr}
\ee
The last term can be identified with the `topological charge' that merely 
depends upon the values of the field on the boundary. The existence of the BPS 
bound also follows from  eq.~(\ref{bog}). A domain wall is the exact solution 
that saturates the BPS  bound and has a non-vanishing topological charge. In 
the case of the double-well scalar potential (\ref{dw}),  eq.~(\ref{bog}) gives
 rise to the well known exact (non-perturbative) BPS solution
\be \lb{dws}
\tilde{\f}= \h \tanh \left( \sqrt{ \fracmm{\l}{2}} \h x \right) \equiv
\h \tanh \left( \fracmm{x}{\d_0} \right)
\ee
whose thickness is thus given by 
\be \lb{thick}
\d_0= \sqrt{\fracmm{2}{\l}}\,\fracmm{1}{\h}
\ee

In the Vilenkin-Linde topological inflation scenario \cite{topo}, should the 
initial conditions for the scalar field $\tilde{\f}$ be randomly distributed 
with large dispersion, a part of the Universe will run to one of the minimums 
of the double-well potential (\ref{dw}) while another part will run to another 
minimum. The domain wall between those parts of the Universe may be thick 
enough, in order to accommodate inflation, when the thickness of the domain 
wall is larger than the Hubble radius of the energy density, ie. \cite{topo}
\be \lb{icon}
\d_0 > H_0^{-1}= \sqrt{ \fracmm{3M^2_{\rm Pl}}{V(0)}} 
\ee
In the case of the double-well potential (\ref{dw}), eqs.~(\ref{thick}) and
(\ref{icon})  yield \cite{topo}
\be \lb{vl}
             \h > {\cal O}\left(\sqrt{6}M_{\rm Pl}\right)
\ee
that should be understood as a rough approximation. The precise condition 
for the existence of a slow-roll inflation with the potential (\ref{dw}) was 
established numerically in ref.~\cite{mae3}:
\be \lb{pre}
             \h > 0.33~\sqrt{8\p}\,M_{\rm Pl}\approx 1.65~M_{\rm Pl}
\ee

The $R^4$-potential (\ref{pot4}) may be roughly approximated by the double-well
 potential (\ref{dw}), as is shown in Fig.~1.~\footnote{That approach
was adopted in ref.~\cite{kal}.} Then the parameter $\h$ is just the distance 
between the minimum and the maximum of our scalar potential, $\h=\f_{\rm max}$.
Then eqs.~({\ref{newvar}) and (\ref{maxv}) imply that in the $R^4$ case 
we have
\be \lb{naiv}
\h \approx 1.345~M_{\rm Pl}
\ee 
while it is even lower for the higher $p>2$. Thus we might conclude that
 a slow-roll inflation is impossible in the $R^{4}$ modified gravity at all. 
However, it is not really the case, because the approximation of the exact 
potential (\ref{pot4}) by the double-well potential (\ref{dw}) is very rough, 
so it cannot be used when the difference between the values of $\h$ in 
eqs.~(\ref{pre}) and (\ref{naiv}) is very small. As a matter of fact, the 
actual scalar potential (\ref{pot4}) does not have an exact $Z_2$ discrete 
symmetry, while its Taylor expansion around the maximum is given by
\be \lb{tay}
\eqalign{
\fracmm{V(\f)}{\frac{2}{9}\cdot 2^{1/3}\cdot V_0}=\fracmm{V(\f)}{V_{\rm max}}
 ~~& = 1-\fracmm{1}{3M^2_{\rm Pl}}(\f-\f_{\rm max})^2
+\fracmm{2\sqrt{2}}{9\sqrt{3}M^3_{\rm Pl}}(\f-\f_{\rm max})^3 \cr
& ~~~- \fracmm{5}{108M^4_{\rm Pl}} (\f-\f_{\rm max})^4 +{\cal O}
\left(\f-\f_{\rm max}\right)^5 \cr}
\ee
where $\f_{\rm max}=\sqrt{3/2}(\ln 3) M_{\rm Pl}$. Since the 3rd derivative is 
non-vanishing and the 4th derivative is negative, the potential (\ref{tay})
cannot be approximated by the double-well potential (\ref{dw}) for any $\l$.
In the next Sec.~4 we study the conditions for a slow-roll inflation with 
the potential (\ref{pot4}).

\section{Slow-roll $R^4$ inflation}

The necessary condition for a slow-roll inflation is the smallness of the 
inflation parameters \cite{llbook},
\be \lb{sroll} \ve(\f)\ll 1 \qquad {\rm and} \qquad \abs{\h(\f)}\ll 1 \ee
where the functions $\ve(\f)$ and $\eta(\f)$ are defined by \cite{llbook}
\be \lb{eta}
\ve(\f) = \fracmm{1}{2} M^2_{\rm Pl} \left( \fracmm{V'}{V}\right)^2\qquad
{\rm and}\qquad 
\h (\f) = M^2_{\rm Pl} \fracmm{V''}{V} 
\ee
and the primes denote the derivatives with respect to the inflaton field $\f$. 
The first condition (\ref{sroll}) implies $\ddt{a}(t)>0$,  whereas the second 
condition (\ref{sroll}) guarantees that inflation lasts long enough, 
via domination of the friction term in the inflaton equation of motion 
(in the slow-roll case):
\be  \lb{fric}
3 H\dt{\f} =- V' \ee
Here $H$ stands for the Hubble function  $H(t)=\dt{a}/a$ in terms of the scale
factor $a(t)$. Equation (\ref{fric}) is to be supplemeted by the Friedmann 
equation 
\be \lb{fried}
 H^2=\fracmm{V}{3M^2_{\rm Pl}}
\ee
It follows from eqs.~(\ref{fric}) and (\ref{fried}) that
\be \lb{feq} 
\dt{\f} =-M_{\rm Pl}\fracmm{V'}{\sqrt{3V}} <0  
\ee
The amount of inflation is measured by the e-foldings number
\be \lb{efol}
N_e = \int^{t_{\rm end}}_t H dt \approx 
\fracmm{1}{M^2_{\rm Pl}} \int^{\f}_{\f_{\rm end}} \fracmm{V}{V'} d\f
\ee
where the $t_{\rm end}$ stands for the (time) end of inflation when one of the
slow-roll parameters becomes equal to $1$. The number of e-foldings  between 
$50$ and $100$ is usually considered to be acceptable.

\begin{figure}[t]
\centering
\includegraphics[width=6cm,clip]{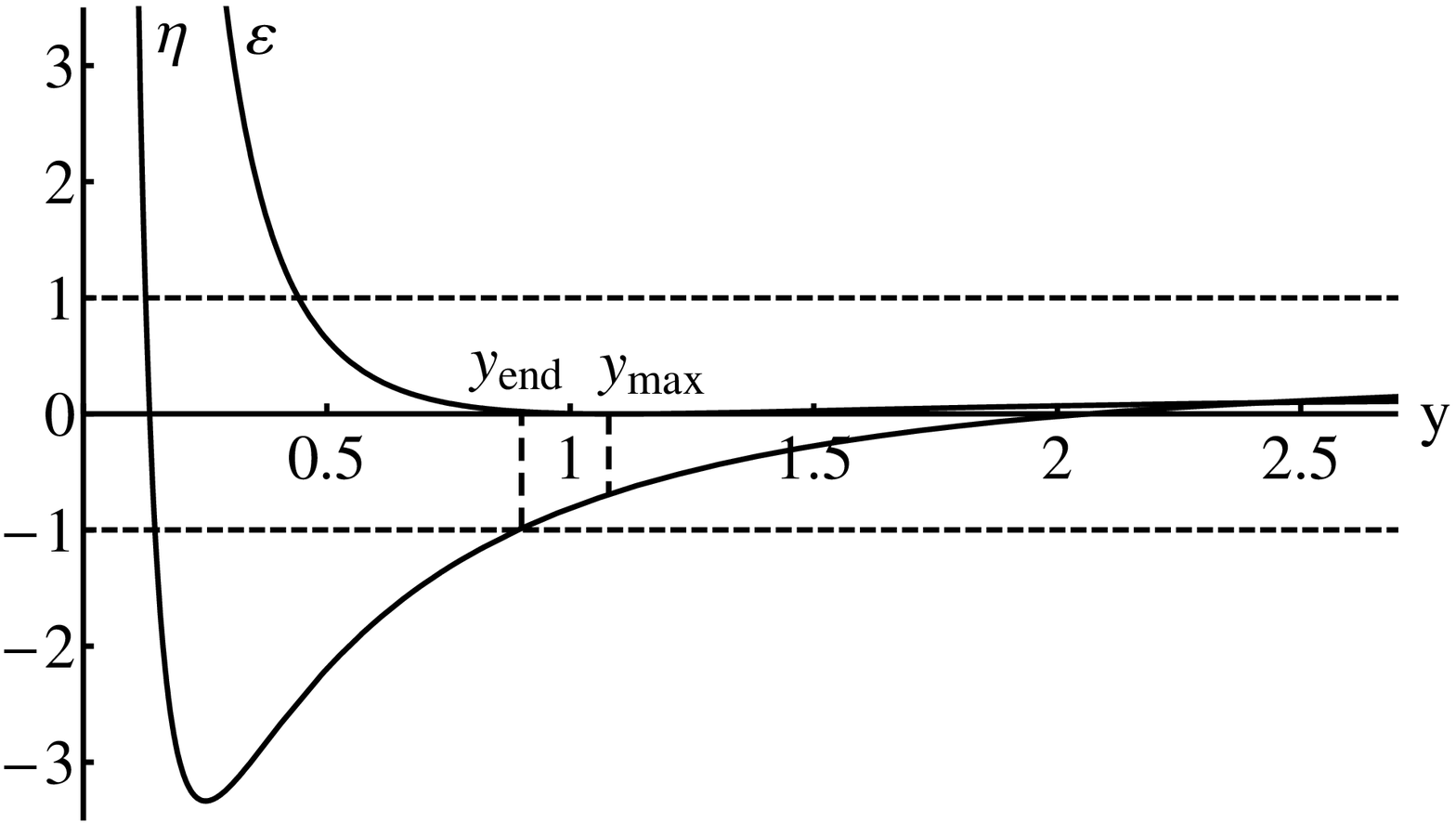}
\caption{\small Graphs of the functions $\ve(y)$ and $\h(\f)$}
\label{fig:3}
\end{figure}

In our case of the scalar potential (\ref{pot4}) we find
\be \lb{epsi}
\ve(y)= \fracmm{1}{3}\left[\fracmm{g'(y)}{g(y}\right]^2=
\fracmm{4}{27}\left[ 1-\fracmm{2}{e^y-1}\right]^2
\ee
and
\be \lb{etae}
\h(y)= \fracmm{2}{3}\left[ \fracmm{g''(y)}{g(y)} \right]=
\fracmm{8}{3(e^y-1)}\left[ -1+\fracmm{e^{2y}}{9(e^y-1)}\right]
\ee
The graphs of those functions are given in Fig.~3.

Inflation is possible when there is a vacuum energy. As is clear from Fig.~1,
it is the case near the maximum $y_{\rm max}=\ln 3\approx 1.1$ . We find
 $ \ve(y_{\rm max})=0$ and $\h(y_{\rm max})=-2/3$, whereas near the minimum 
$y=0^+$ both $\ve(0^+)$ and $\h(0^+)$ diverge to infinity. The equation 
$\ve(y)=1$ has two solutions at $e^y=1.55$ and $e^y=0.44$, whereas the equation
 $\h(y)=1$ also has two solutions at $e^y=2.44$ and $e^y=1.15$. Therefore, it
is clear that a slow-roll inflation from the maximum towards the minimum of the
 potential has an exit at  $y_{\rm end}\approx\ln 2.44\approx 0.89$, whereas 
the runaway inflation from the maximum towards $y=+\infty$ never ends.
  
As regards the e-foldings number in the case of the slow-roll inflation towards
the minimum, for $(I):~y_{\rm end}<y<y_{\rm max}~$, we find 
\be \lb{efol2}
\eqalign{
N^I_e(y)  & ~~~=\fracmm{3}{2} \int^{y}_{y\low{\rm end}} \fracmm{g(y)}{g'(y)} dy
=\fracmm{9}{4}\int^{y\low{\rm end}}_{y} \left(\fracmm{e^y-1}{3-e^y}\right) 
dy \cr
& ~~~=\fracmm{3}{2}\ln\left[ \fracmm{-1+3e^{-y\low{\rm end}}}{-1+3e^{-y}}
\right]-\fracmm{9}{4}(y-y\low{\rm end})
\approx \fracmm{3}{2}\ln\left[ \fracmm{e^y}{3-e^y}\right] -\fracmm{9}{4}y -0.2
\cr}
\ee
Similarly, for the runaway inflation $(II):~y_{\rm max}<y<y\low{\rm large}$ 
we find
\be \lb{efol3}
N^{II}_e(y) = \fracmm{9}{4}\int^{y\low{\rm large}}_{y} \left(
\fracmm{e^y-1}{e^y-3}\right) dy 
\ee

When $y_{\rm large}\to+\infty$, the slow-roll parameters approach finite 
values, $\ve\to 4/27$ and $\h\to 8/27$, whereas  $N_e^{II}$ linearly diverges, 
as expected (see Fig.~3).

It is straghtforward to get explicit solutions to the slow-roll equations of 
motion (\ref{fried}) and (\ref{feq}) by using the exact potential 
(\ref{pot4}), though those solutions are not very illuminating. So, we restrict
 ourselves to presenting only the leading terms, in the vicinity of the 
maximum, {\it viz.}
\be \lb{fsol}
\f(t)\approx \f_{\rm max} + (\f_{\rm ini}-\f_{\rm max})
\exp\left[ \fracmm{m^2}{3H_0}(t-t_{\rm ini})\right]
\ee
and
\be \lb{scalef}
a(t)\propto e^{H_0t} \left( 1- \fracmm{ (\f_{\rm ini}
-\f_{\rm max})^2}{16M^2_{\rm Pl}}
\exp\left[ \fracmm{2m^2}{3H_0}(t-t_{\rm ini})\right]\right)
\ee
where we have introduced the Hubble constant as usual, ie.
\be \lb{newnot}
H_0=\fracmm{1}{M_{\rm Pl}} \sqrt{ \fracmm{V_{\rm max}}{3}} 
\qquad {\rm with}\qquad V_{\rm max}=V(y_{\rm max}) = 
\fracmm{M^2_{\rm Pl}M^2}{12\cdot 2^{1/3}}
\ee 
and the mass parameter
\be \lb{mass}
m^2=\fracmm{2V_{\rm max}}{3M^2_{\rm Pl}}=
\fracmm{M^2}{18\cdot 2^{1/3}}
\ee
in agreement with the Taylor expansion (\ref{tay}) of the inflaton potential.

We are now ready to confront the $R^4$ model with cosmological observations. 
The first relevant physical observable is given by the amplitude of the initial
 perturbations, $\D^2_R=M^4_{\rm Pl}V/(24\p^2\ve)$. Equating its theoretical
and experimental values yields \cite{llbook} 
\be \lb{ampl}
\left. \left(\fracmm{V}{\ve}\right)^{1/4}\right|_{y=y_{\rm end}} =
0.027\,M_{\rm Pl}=
6.6\cdot 10^{16}~{\rm GeV}
\ee
This equation determines the normalization of the $R^4$-term in the action. By
using $g(y_{\rm end})\approx 0.27$ and $\ve(y_{\rm end})\approx 0.022$ we find 
 \be \lb{estim}
M\approx 4.3\cdot 10^{-4}M_{\rm Pl}
\ee
However, the key test is provided by the observed value 
of the spectral index $n_s$, as given by the recent WMAP5 data 
\cite{wmap5},
\be \lb{sind5} 
 n_s = 0.960 \pm 0.013  \ee
The theoretical value of $n_s$ is given by \cite{llbook}
\be \lb{sind}
n_s  = 1+2\h -6\ve = 1+ \fracmm{16}{3(e^y-1)}\left[ -1+
\fracmm{e^{2y}}{9(e^y-1)}
\right] -\fracmm{8}{9}\left[-1+\fracmm{2}{e^y-1}\right]^2
 \ee
where we have used eqs.~(\ref{epsi}) and (\ref{etae}) in our case. 
Unfortunately, as regards the relevant values of $y$ during the slow-roll
inflation towards the minimum, the $\ve(y)$ is always less than $0.03$ 
(so it does not play a significant role here), whereas the value of $\h(y)$ 
always belongs to the interval $2/3\leq \abs{\h}\leq 1$, so that the $\h$ is 
unacceptably large during inflation ($\h$-problem). One arrives at the same 
conclusion when using another standard equation for the spectral 
index,~\footnote{See eg., ref.~\cite{kal} and references therein.}
\be \lb {sind2}
n_s= 1-\fracmm{2m^2}{3H^2_0} \ee
whose value at the maximum is given by $n_s(y_{\rm max}) =1-4/3=-1/3$. 
Similarly, when considering the runaway inflation, we find that 
$n_s\to 19/27$. 

\section{Conclusion}

In conclusion, a slow-roll inflation with an exit is possible in the modified
$R^4$-gravity model, but it does not survive the observational tests because 
of the $\h$-problem. This qualitative conclusion agrees with that of 
ref.~\cite{kal} though our quantitative results are different. 

Two possible cures of the $\h$-problem were already proposed in 
ref.~\cite{kal}. As regards (1) adding the $R^2$ term to the $R^4$ term 
\cite{kal}, it would be against our basic (or minimal) motivation in Sec.~1 
(see, however, ref.~\cite{zhuk}). Of course, the $R^2$ term may also be 
generated in the process of superstring compactification, however, we wanted to
check just the topological inflation. As regards (2) taking into account the 
renormalization of the matter stress-energy tensor 
\cite{star,kal}, it may lead to sufficient inflation with an acceptable value 
of $n_s$, but without the predictive power because of the uncertainty in the 
number of relevant string degrees of freedom.
 
Of course, it may be not difficult to find a function $f(R)$ with the shape
similar to that in Fig.~1, whose $y_{\rm max}$ and $V_{\rm max}$ would have the
desired values for topological inflation with the desirable value of $n_s$.
However, we are not aware of any string theory output that would constrain that
function in any way. One of the natural first steps would be the embedding of 
topological inflation into modified supergravity  \cite{gket}.

\section*{Acknowledgements}

The authors are grateful to N. Iizuka, N. Kaloper, A. Starobinsky and A. Zhuk 
for correspondence.
\vglue.2in

\end{document}